\documentclass[
reprint,
superscriptaddress,
bibnotes,
amsmath,amssymb,
aps,
pra,
]{revtex4-1}

\usepackage{graphicx,import}
\usepackage{dcolumn}
\usepackage{braket}
\usepackage{textcomp, gensymb}
\usepackage{bm}
\usepackage{amsmath}
\usepackage{float}
\usepackage{mathtools}
\usepackage{color}
\usepackage[dvipsnames,table,xcdraw]{xcolor}
\usepackage[utf8]{inputenc}
\usepackage{csquotes}
\usepackage{thmtools}

\usepackage{dsfont}
\usepackage{bbm}

\usepackage{hyperref}

\usepackage[normalem]{ulem}

\usepackage{outline}
\usepackage{physics}
\usepackage{multirow}
\usepackage{siunitx}
\usepackage{tikz}
\usetikzlibrary{shapes.geometric, arrows,positioning}
\usepackage{qcircuit}
\usepackage{lipsum}
\usepackage{chemfig}
\usepackage{silence}
\usepackage{algorithm}
\usepackage{algpseudocode}

\makeatletter
\def\p@subsection{}
\makeatother
\makeatletter
\def\p@subsubsection{}
\makeatother

\allowdisplaybreaks

\begin{document}

\title{Robust spin-qubit control in a natural Si-MOS quantum dot using phase modulation}
\author{Takuma Kuno}
\affiliation{Research and Development Group, Hitachi, Ltd., Kokubunji, Tokyo 185-8601, Japan}
\affiliation{Department of Electrical and Electronic Engineering, Institute of Science Tokyo, Meguro, Tokyo 152-8552, Japan}

\author{Takeru Utsugi}
\affiliation{Research and Development Group, Hitachi, Ltd., Kokubunji, Tokyo 185-8601, Japan}

\author{Andrew J. Ramsay}
\affiliation{Hitachi Cambridge Laboratory, J. J. Thomson Ave., Cambridge, CB3
    0HE, United Kingdom}

\author{Normann Mertig}
\affiliation{Hitachi Cambridge Laboratory, J. J. Thomson Ave., Cambridge, CB3
    0HE, United Kingdom}

\author{Noriyuki Lee}
\affiliation{Research and Development Group, Hitachi, Ltd., Kokubunji, Tokyo 185-8601, Japan}

\author{Itaru Yanagi}
\affiliation{Research and Development Group, Hitachi, Ltd., Kokubunji, Tokyo 185-8601, Japan}

\author{Toshiyuki Mine}
\affiliation{Research and Development Group, Hitachi, Ltd., Kokubunji, Tokyo 185-8601, Japan}

\author{Nobuhiro Kusuno}
\affiliation{Research and Development Group, Hitachi, Ltd., Kokubunji, Tokyo 185-8601, Japan}

\author{Raisei Mizokuchi}
\affiliation{Department of Electrical and Electronic Engineering, Institute of Science Tokyo, Meguro, Tokyo 152-8552, Japan}

\author{Takashi Nakajima}
\affiliation{Center for Emergent Matter Science, RIKEN, Wako‑shi, Saitama 351‑0198, Japan}

\author{Shinichi Saito}
\affiliation{Research and Development Group, Hitachi, Ltd., Kokubunji, Tokyo 185-8601, Japan}

\author{Digh Hisamoto}
\affiliation{Research and Development Group, Hitachi, Ltd., Kokubunji, Tokyo 185-8601, Japan}

\author{Ryuta Tsuchiya}
\affiliation{Research and Development Group, Hitachi, Ltd., Kokubunji, Tokyo 185-8601, Japan}

\author{Jun Yoneda}
\affiliation{Academy of Super Smart Society, Institute of Science Tokyo, Meguro, Tokyo 152-8552, Japan}
\affiliation{Department of Advanced Materials Science, University of Tokyo, Kashiwa, Chiba 277-8561, Japan}

\author{Tetsuo Kodera}
\affiliation{Department of Electrical and Electronic Engineering, Institute of Science Tokyo, Meguro, Tokyo 152-8552, Japan}

\author{Hiroyuki Mizuno}
\affiliation{Research and Development Group, Hitachi, Ltd., Kokubunji, Tokyo 185-8601, Japan}

\date{\today}

\begin{abstract}
    \section*{abstract}
    Silicon quantum dots are one of the most promising candidates for practical quantum computers because of their scalability and compatibility with the well-established complementary metal-oxide-semiconductor technology.
    However, the coherence time is limited in industry-standard natural silicon because of the $^{29}$Si isotopes, which have non-zero nuclear spin.
    Here, we protect an isotopically natural silicon metal-oxide-semiconductor (Si-MOS) quantum dot spin qubit from environmental noise via electron spin resonance with a phase-modulated microwave (MW) drive.
    This concatenated continuous drive (CCD) method extends the decay time of Rabi oscillations from 1.2 $\mathrm{\mu s}$ to over 200 $\mathrm{\mu s}$.
    Furthermore, we define a protected qubit basis and propose robust gate operations.
    We find the coherence time measured by Ramsey sequence is improved from 143 ns to 40.7 $\mu$s compared to that of the bare spin qubit.
    The single qubit gate fidelity measured with randomized benchmarking is improved from 95\% to 99\%, underscoring the effectiveness of the CCD method.
    The method shows promise for improving control fidelity of noisy qubits, overcoming the qubit variability for global control, and maintaining qubit coherence while idling.

\end{abstract}

\maketitle

\noindent\texttt{takuma.kuno.pg@hitachi.com}

\section{Introduction}
\label{sec:introduction}

Silicon quantum dots are promising quantum computing platforms because of their small size, kelvin-scale temperature operation~\cite{petit2020universal,yang2020operation,huang2024high,undseth2023hotter} and compatibility with the well-established complementary metal-oxide-semiconductor technology~\cite{gonzalez2021scaling,li2018crossbar,le2020low,curry2019single,7838410,de2023silicon,zwerver2022qubits}.
The implementation of fault-tolerant quantum computation requires high operational fidelity and long coherence times, and the key to achieving this is the realization of a qubit that is robust against environmental noise.
For spin qubits in silicon quantum dots, the main sources of noise are the nuclear spins of $^{29}$Si isotopes~\cite{khaetskii2002electron,assali2011hyperfine} and charge noise acting via the spin-orbit interaction~\cite{yoneda2018quantum,bermeister2014charge,huang2014electron,freeman2016comparison}.
The noise can be reduced by device engineering approaches such as isotopic purification \cite{witzel2010electron,muhonen2014storing,itoh2014isotope,acharya2024highly} and optimization of oxide interfaces~\cite{elsayed2022low}.
These noise mitigation efforts have enabled remarkably high gate fidelities in silicon-based spin qubits~\cite{veldhorst2014addressable,huang2019fidelity,xue2022quantum,noiri2022fast,steinacker2025industry,wu2025simultaneous}. However, many of these studies rely on frequent feedback and calibration~\cite{huang2019fidelity,noiri2022fast,steinacker2025industry,wu2025simultaneous}. As we scale up to multi-qubit systems, the overhead of this approach may become more challenging. Therefore, the design of noise-resilient gate operations is also a valuable strategy. One promising approach is pulse design.
Good pulse design considers the noise characteristics of the qubit~\cite{yang2019silicon,rimbach2023simple}, which for spin qubits is dominated by phase noise.
Hence, the operation with the lowest fidelity, which is dominated by the phase noise, is the identity or idle with no applied microwave (MW) drive, that is, to simply wait.
This is because application of a MW drive causes the spin to rotate, averaging out the influence of low frequency noise, and dynamically decoupling the spin from the environment~\cite{laucht2017dressed}.
As a result, the observed coherence time depends on the noise filter applied by the pulse sequence~\cite{cywinski2008enhance}.

A good control scheme incorporates dynamic decoupling.
Traditionally, pulse-sequences, such as Hahn echo~\cite{hahn1950spin} and CPMG~\cite{zhang2014protected,van2012decoherence} are used for idle operations.
However, as the number of refocusing pulses increases, the coherence is often lost due to pulse-area errors~\cite{wang2012effect}.
An alternative approach uses a modulated continuous drive to dynamically decouple the spins.
This is an attractive proposition, since the dynamic decoupling is always on, even during operations.
One approach explored for silicon quantum dots is the sinusoidally modulated, always rotating and tailored (SMART) protocol~\cite{hansen2021pulse,hansen2022implementation}, where the alternating phase of a sinusoidally amplitude modulated drive continually refocuses the spin against detuning errors.
The SMART protocol has recently been used for an entangling operation~\cite{hansen2024entangling}.
Another protocol explored in the spin defect literature is the concatenated continuous drive (CCD) protocol, first developed for NV-centers~\cite{wang2020coherence,stark2017narrow}.
This CCD protocol has proven effective at decoupling an electron spin from nuclear spin noise in hexagonal boron nitride, a III-V material~\cite{ramsay2023coherence} and a silicon quantum dot~\cite{kuno2024concatenated}.
In the CCD protocol, the qubit can be protected using only phase modulation~\cite{cohen2017continuous}, which avoids potential issues associated with amplitude modulation, such as nonlinearity in MW amplifiers or I/Q mixers, and qubit instability caused by time-dependent heating due to the amplitude modulation itself.
A similar phase-modulation scheme has been explored for silicon hole spins, however the experiments used a transport measurement, limiting the ability to fully evaluate the potential for coherence protection~\cite{bosco2023phasehole}.
The CCD method has been shown to be extremely useful for improving robustness, but so far there has been no verification of its effectiveness in the gate fidelity, and in particular, a suitable control scheme for silicon spin qubits has not yet been elucidated.

Here, we present a phase-modulation protocol for control of a single spin qubit in a Si-MOS quantum dot fabricated on a fully depleted silicon on insulator (FD-SOI) substrate.
The electron spin exhibits coherence times of $T_2^*=143$ ns and Rabi decay time of $T_2^{\rm{Rabi}}=1.2$ $\mu$s, while spin echo experiment yields a significantly longer coherence time of $T_2^{\rm{echo}}=230$~$\mu$s. This property suggests that coherence is limited by low-frequency noise, attributed to nuclear spins from $^{29}$Si atoms and charge noise, as previously reported in natural silicon devices~\cite{hanson2007spins,kawakami2014electrical}.
Using a CCD drive, the spin can be continuously dynamically decoupled from the environment, extending the coherence time of Rabi oscillations from 1.2 $\mathrm{\mu s}$ to over 200 $\mathrm{\mu s}$~\cite{kuno2024concatenated}.
To exploit this enhanced coherence time, we define a protected qubit in a double dressed basis and then propose and demonstrate a method for two-axis control and readout of the protected qubit.
We find the quality factor of spin rotations is improved from 2.2 to 25.0 by the proposed CCD method.
We further define a set of single qubit Clifford gates and perform randomized benchmarking experiment~\cite{knill2008randomized,muhonen2015quantifying}.
The average gate fidelity is improved from 95\% to $99.1\pm0.1$\%, which is at the edge of the threshold for surface code error correction~\cite{fowler2012surface}, despite the use of a natural Si-MOS device.
In previous studies, even isotopically purified $^{28}$Si devices exhibit detuning fluctuations that require the use of feedback to continually update the qubit calibration~\cite{dumoulin2024silicon,huang2019fidelity,steinacker2025industry}. By contrast, CCD achieves high-fidelity operations that are intrinsically robust, and can be used without active feedback, simplifying the control.

\section{Results}
\subsection{Device}
\begin{figure*}[htbp]
    \includegraphics[width=\columnwidth*2]{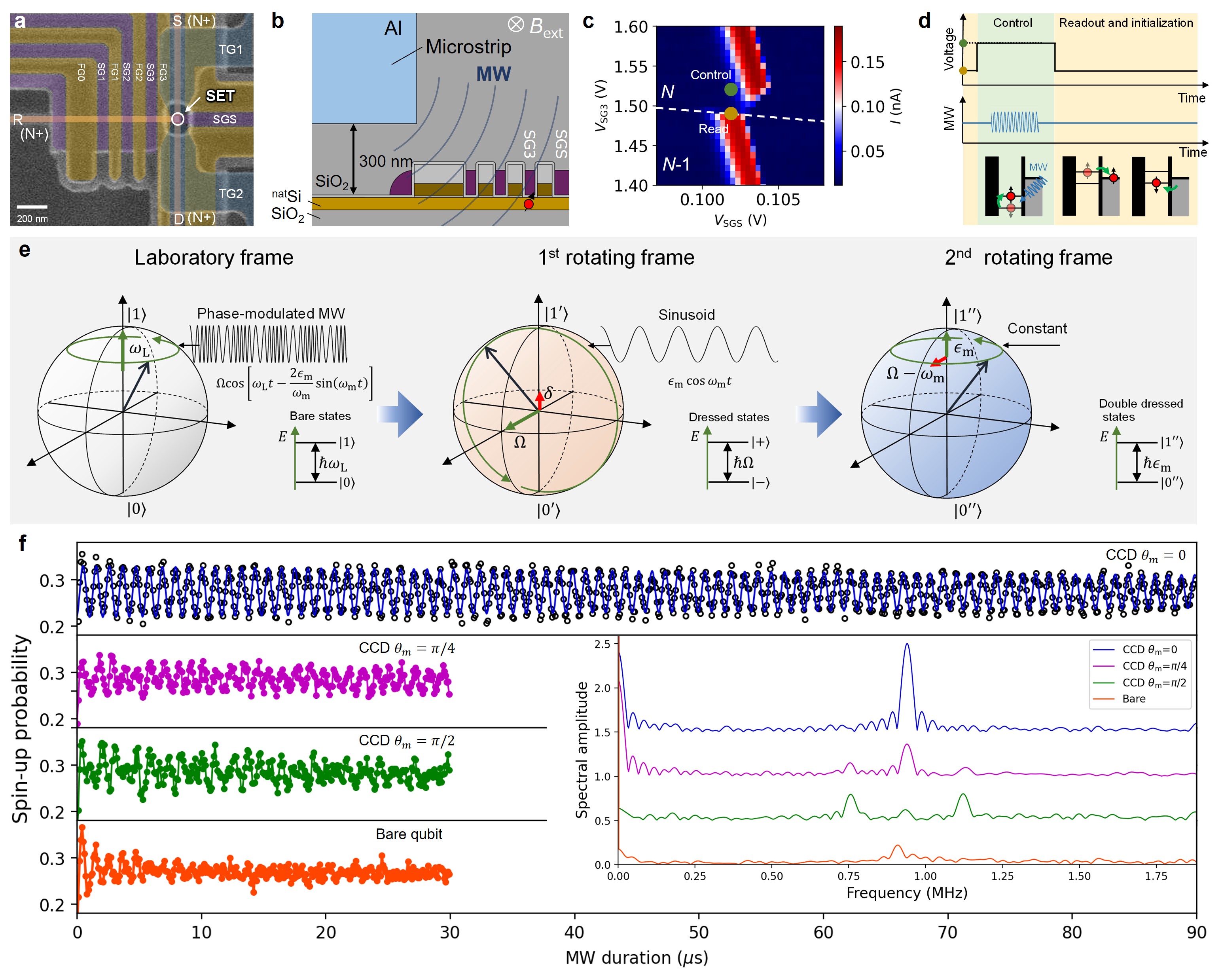}
    \caption{{\bf CCD-protected spin qubit in a natural Si-MOS quantum dot}. {\bf a} False-color scanning electron microscope image of the device.
    The device is designed to host an array of three quantum dots under gates SG1, SG2, and SG3 in a Si/SiO2 channel along with an SET.
    We manipulate the electron spin under gate SG3.
        {\bf b} Schematic of the Si-MOS device. An external magnetic field $B_{\rm{ext}}= 825$ mT is applied to the device.
        {\bf c} Charge stability diagram measured as a function of the gate voltages $V_{\rm{SGS}}$ and $V_{\rm{SG3}}$. The white dashed line shows the charge transition line and $N$ denotes the number of electrons in the quantum dot under the SG3 gate.
    Colored dots show the SG3 and the SGS gate biases for control and readout of the spin state.
        {\bf d} Gate-voltage pulse cycle for energy-selective initialization and readout.
        {\bf e} Principle of coherence protection by phase-modulated MW. In the laboratory frame, the bare qubit is protected from spin-flips by an energy-gap $\hbar\omega_{\rm{L}}$, but detuning errors in $\omega_{\rm{L}}$ can cause phase-drift. In the first rotating frame of a resonant MW drive, the qubit is protected from detuning errors due to an energy-gap $\hbar\Omega$ between the dressed states, and the phase-modulation appears as an a.c. drive. In the second rotating frame with the phase-modulation, the qubit is protected from fluctuations in the MW-drive by an energy-gap $\hbar\epsilon_{\rm{m}}$, and the MW-drive appears as a d.c. B-field. In other words, the qubit is dynamically decoupled from noise by a rotation about two axes.
        {\bf f} Comparison of the bare qubit Rabi oscillations and the CCD-protected Rabi oscillations in the case with $\theta_{\rm{m}}=0, \pi/2$, and $\pi/4$. For $\theta_{\rm{m}}=0$, we use the same data as ref~\cite{kuno2024concatenated}.
    The inset shows the Fourier transform spectrum of these Rabi oscillations. When we set $\theta_{\rm{m}}=\pi/4$, we can observe the Mollow triplet. When $\theta_{\rm{m}}=0$, which we call the idle pulse state, only the center peak appears, corresponding to the stabilized Rabi oscillations.
    }\label{FIG1}
\end{figure*}

The device is fabricated in Hitachi's R$\&$D clean room using standard CMOS process.
A false-colored scanning electron microscope image and a cross-sectional schematic of the device, a 1x3 quantum dot array with a single electron transistor (SET), are shown in Figs.~\ref{FIG1}a and b.
The spin qubit is located under gate SG3.
Energy selective preparation and readout of the spin is used~\cite{elzerman2004single} (Fig.~\ref{FIG1}d), and the electron spin resonance (ESR) drive field is applied with an Aluminium microstrip.
All measurements are taken at $B_{\rm{ext}}$ = 825 mT, where the resonance frequency is 22.567 GHz.

The bare qubit properties are described in detail in Supplementary Note 1.
The qubit decay time $T_2^{\rm{Rabi}}$ in the natural silicon channel is about microseconds, mainly due to low-frequency noise arising from charge noise and nuclear spins of $^{29}$Si isotopes.
On the other hand, we observe a relatively long spin echo time $T_2^{\rm{echo}}$ of 230 $\mu$s.
This suggests sources of high frequency noise (a few kHz) in our Si-MOS channel fabricated by the shallow trench isolation (STI) process are weak.
However, the gate fidelity is limited by the $T_{\rm{2}}^{\rm{Rabi}}$ time, which is dominated by low frequency noise.
Hence, spin control robust against low frequency noises is desired.

\subsection{Coherence protection}

One approach to protect a qubit from noise is to dress its states by applying an always-on MW drive \cite{laucht2017dressed}.
This defines a dressed or protected rotating qubit basis, where the qubit is dynamically decoupled from the noise by averaging out the low frequency noise.
In the CCD scheme, by applying the amplitude or phase modulated MW, the qubit is protected in the double dressed state against both detuning and the Rabi frequency errors, resulting in a longer coherence time (Fig.~\ref{FIG1}e).
In our experiment, we implement CCD protection using a phase modulated MW drive.
This protocol is expected to be more stable than modulating the MW drive amplitude, which is affected not only by the stability of the MW source itself, but also by variations in the $g$-factor.
The system Hamiltonian of the phase-modulated spin in the laboratory frame is written as
\begin{align}
     & H_{\rm{lab}}(t)=\frac{\hbar(\omega_{\rm{mw}}+\delta)}{2} \sigma_z \nonumber                                                                                                \\
     & + \hbar\Omega \cos\left[\omega_{\rm{mw}}t+\phi_{\rm{mw}} - \frac{2\epsilon_{\rm{m}}}{\omega_{\rm{m}}}\sin(\omega_{\rm{m}}t-\theta_{\rm{m}})\right]\sigma_x \label{eq-001},
\end{align}
where $\sigma_i (i\in{x,y,z})$ are the Pauli matrices, $(\omega_{\rm{mw}}+\delta)/2\pi = \omega_{\rm{L}}/2\pi$ is the
Larmor frequency, $\delta/2\pi$ is the detuning error, $\Omega/2\pi$
is the Rabi frequency, and $\omega_{\rm{mw}}/2\pi$ and
$\phi_{\rm{mw}}$ are the MW frequency and phase, respectively.
$\epsilon_{\rm{m}}$, $\omega_{\rm{m}}$ and $\theta_{\rm{m}}$ are the modulation parameters of the CCD protocol.
The first term is the Zeeman Hamiltonian and the second term is the phase-modulated MW-drive.
The modulation strength, denoted by $\epsilon_{\rm{m}}$, is the gate speed of the CCD-protected qubit and should be smaller than $\Omega$ to cancel the counter-rotating term as discussed in the next section.

We introduce the first rotating frame defined by $H_0^{(1)} = \frac{\hbar}{2}\frac{d}{dt}\Phi_{\rm{CCD}}\sigma_z$, where $\Phi_{\rm{CCD}} = \omega_{\rm{mw}}t-\frac{2\epsilon_{\rm{m}}}{\omega_{\rm{m}}}\sin(\omega_{\rm{m}}t-\theta_{\rm{m}})$.
The Hamiltonian is described as,
\begin{align}
     & H_{\rm{rot}}^{(1)}(t)=\frac{\hbar\delta}{2} \sigma_z +\frac{\hbar\Omega}{2}\sigma_{\phi_{\rm{mw}}} + \hbar\epsilon_{\rm{m}}\cos(\omega_{\rm{m}}t-\theta_{\rm{m}})\sigma_{z}, \label{eq-002}
\end{align}
where ${\sigma_{\phi_{\rm{mw}}}=\cos(\phi_{\rm{mw}})\sigma_x + \sin(\phi_{\rm{mw}})\sigma_y}$ and we omit the counter-rotating term.
In the first rotating frame, the usual Rabi drive field (second term of Eq. \eqref{eq-002}) is perpendicular to the detuning error, which suppresses the phase shift since the detuning error term only slightly tilts the Rabi drive field.

We now move to the second rotating frame, where a qubit defined by the double dressed states is protected against the Rabi frequency error.
Taking the second rotating frame defined by $H_0^{(2)}=\frac{\hbar\omega_{\rm{m}}}{2}\sigma_{\phi_{\rm{mw}}}$, we obtain
\begin{align}
     & H_{\rm{rot}}^{(2)}(t)=\frac{\hbar(\Omega-\omega_{\rm{m}})}{2}\sigma_{\phi_{\rm{mw}}}                                                                      \notag \\
     & +\frac{\hbar\epsilon_{\rm{m}}}{2} \left[ \cos(\theta_{\rm{m}})\sigma_z+\sin(\theta_{\rm{m}})\sigma_{\phi_{\rm{mw}+\frac{\pi}{2}}}  \right], \label{eq-003}
\end{align}
where we set $\delta = 0$ and omit the counter-rotating term.
When we employ the on-resonance condition in the second rotating frame, where $\omega_{\rm{m}}$ is set to the mean value of $\Omega$, the first term in Eq. \eqref{eq-003} is considered to be the Rabi frequency error caused by, for example, the MW instability and the variations in the $g$-factor.
This Rabi frequency error is perpendicular to the CCD drive field of second term in Eq. \eqref{eq-003}, reducing the effect of the Rabi drive error.
In essence, for the CCD protocol, the qubit is protected from both the detuning and the Rabi frequency error by generating the energy gap of $\hbar\Omega$ and $\hbar\epsilon_{\rm{m}}$, respectively.
For more details on the theory, see Supplementary Note 2 and Supplementary Note 3.

When we set $\omega_{\rm{m}} = \Omega$ and $\theta_{\rm{m}}=0$, which we call the idle pulse state, the Hamiltonian in the second rotating frame Eq. (3) reduces to $H_{\rm{rot}}^{(2)}(t)=\frac{\hbar\epsilon_{\rm{m}}\sigma_z}{2}$.
Since the idle pulse does not change the $z$ component of the qubit and acts as a periodic refocusing of the spin to maintain coherence, measuring the spin-up probability in the laboratory frame reveals the prolonged Rabi oscillations.
Figure~\ref{FIG1}f compares the bare qubit and the CCD-protected Rabi oscillations under different modulation conditions for $\omega_{\rm{m}}=\Omega = 2\pi\times940$ kHz, and $\epsilon_{\rm{m}} = \omega_{\rm{m}}/5$.
When $\theta_{\rm{m}}=0$, the CCD-protected Rabi oscillations continue significantly longer in contrast to the bare qubit, which decays rapidly.
We extract the CCD-protected Rabi decay time $T_{\rm{CCD-idle}}$ as 211 $\mu$s.
This is more than 100 times longer than the bare qubit decay time $T_2^{\rm{Rabi}}$ of 1.2 $\mu$s.
When we change the modulation phase $\theta_{\rm{m}}$ to $\pi/4$, we see a strong beating pattern.
As seen in the Fourier domain (see the inset of Fig.~\ref{FIG1}f), this beating pattern consists of a Mollow triplet with center frequency $\omega_{\rm{m}}$ and sidebands at $\omega_{\rm{m}}\pm\epsilon_{\rm{m}}$, as previously reported in spin defects~\cite{ramsay2023coherence,wang2021observation}.
The modulation phase, $\theta_{\rm{m}}$, controls the relative magnitude of the sidebands.
As discussed in the next section, this knob can be used to control a dressed qubit protected against the noise generated by charge noise and nuclear spins of $^{29}$Si isotopes.

\subsection{Control and readout of a CCD-protected qubit}

\begin{figure}[htbp]
    \includegraphics[width=\columnwidth]{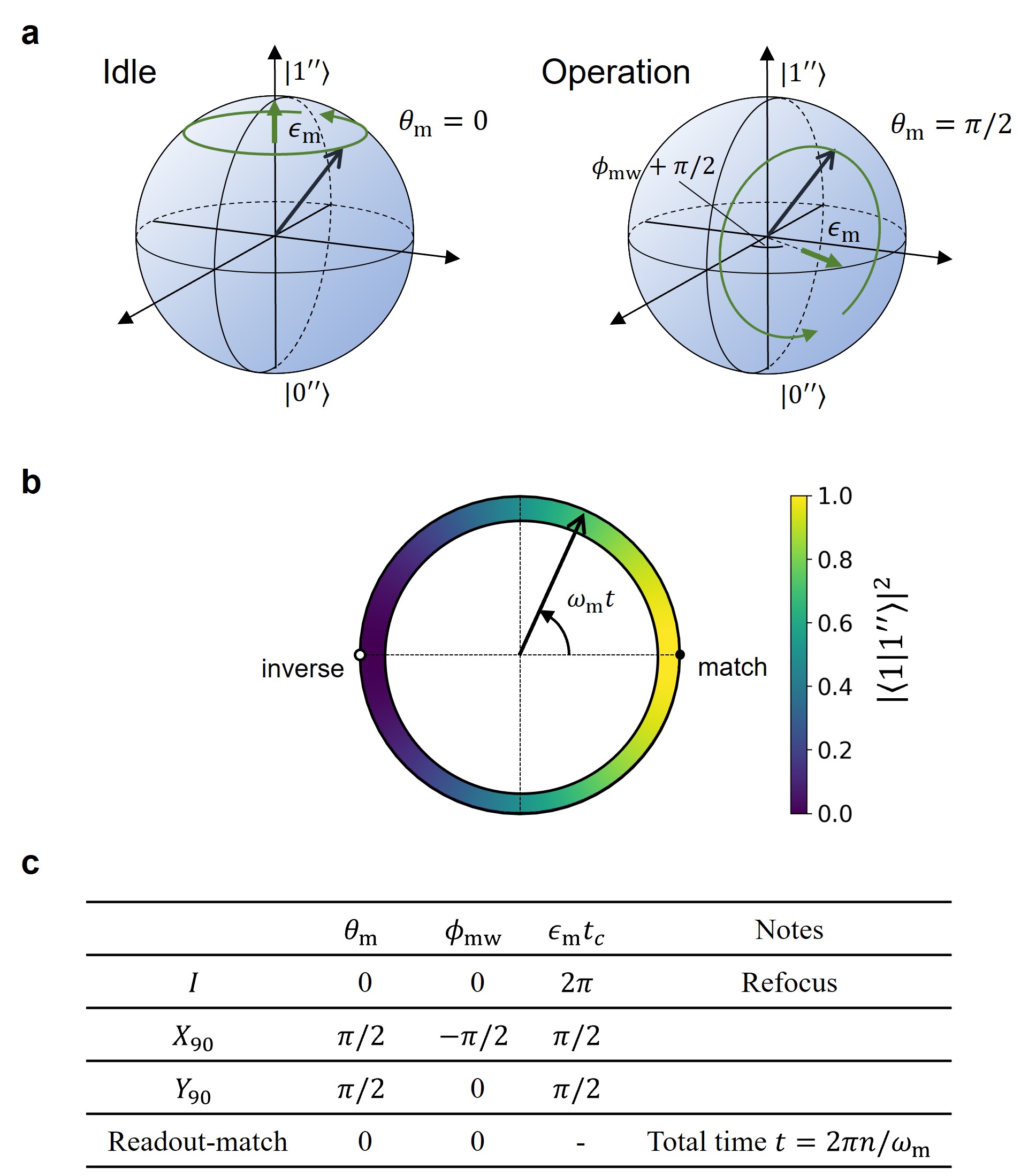}
    \caption{{\bf Protected qubit control and readout protocol.}
        {\bf a} Schematic of the Bloch sphere along with the Hamiltonian vector for idle and operation. When executing idle, $\theta_{\rm{m}}$ is set to 0 and the Hamiltonian vector aligned to the $z$-axis. The idle pulse is used to maintain coherence between operations and after the operation pulses to match the laboratory and the second rotating frame bases for readout.
    To perform gate operations, we set $\theta_{\rm{m}}=\pi/2$ and change the azimuth angle of the Hamiltonian vector in the $x$-$y$ plane, enabling spin rotation around an arbitrary axis perpendicular to the $z$-axis.
    {\bf b} Indicator representing the periodic readout matching between the laboratory and second rotating frames. The color of the ring shows the inner product of the laboratory frame basis state and the second rotating frame basis state, $\left\lvert \left\langle 1 \lvert  1'' \right\rangle  \right\rvert^2$. To achieve readout matching the total pulse length satisfies $\omega_{\rm{m}} t=2\pi n$.
        {\bf c} Example drive conditions for diffrent single protected qubit operations. $t_c$ is the pulse duration.
    }
    \label{FIG2}
\end{figure}

\begin{figure*}
    \includegraphics[width=\columnwidth*2]{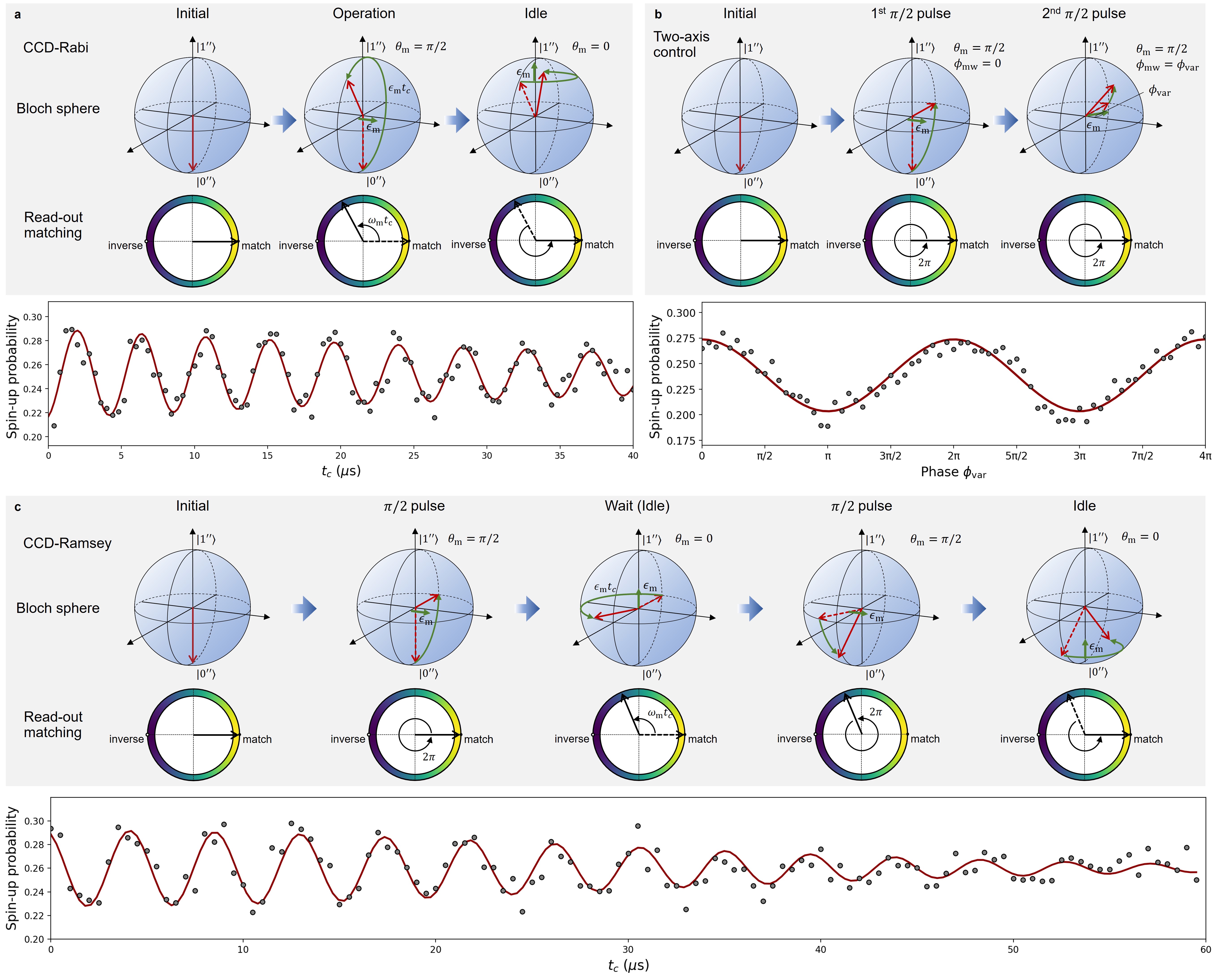}
    \caption{{\bf Demonstration of manipulation and readout of the CCD-protected qubit.}
        All measurements are conducted under $\epsilon_{\rm{m}} = \omega_{\rm{m}}/4 =2\pi\times235$ kHz.
        The top panel of each figure shows time evolutions of the CCD-protected qubit using the Bloch sphere representation of the effective field $\epsilon_{\rm{m}}$ and the readout indicators.
            {\bf a} CCD-Rabi experiment. After initialization, the operation pulse of the duratoin of $t_c$ is applied, followed by the idle pulse for readout matching.
            {\bf b} Two-axis control experiment. Two consecutive $\pi/2$ pulses, with MW phases 0 and $\phi_{\rm{var}}$, are applied. The measured spin-up probability as a function of $\phi_{\rm{var}}$ oscillates with a period of $2\pi$, demonstrating the control of the rotation axis in the CCD qubit basis.
            {\bf c} CCD-Ramsey experiment. Two $\pi/2$ pulses and an idle pulse between them are applied. The spin-up probability is plotted as a function of the duration of the idle pulse $t_c$, and the coherence time is evaluated from the decay.
    }
    \label{FIG3}
\end{figure*}

To exploit the advantage of coherence protection via the CCD protocol for qubit applications, we need to define, control and readout the CCD-protected qubit states.
In the protected frame, for the on-resonance condition drive, the Hamiltonian reads:
\begin{align}
     & H_{\rm{rot}}^{(2)}=
    \frac{\hbar\epsilon_{\rm{m}}}{2} \left\{ \cos(\theta_{\rm{m}})\sigma_z+\sin(\theta_{\rm{m}})\sigma_{\phi_{\rm{mw}}+\frac{\pi}{2}} \right\} \label{eq-004}.
\end{align}
We start by defining an idle-pulse by MW drive with $\theta_{\rm{m}}=\phi_{\rm{m}}=0$, as demonstrated in Fig. \ref{FIG1}f , for a duration of $2\pi/\epsilon_{\rm{m}}$.
The idle pulse defines the protected qubit basis as $\{ \vert 0''\rangle$, $\vert 1''\rangle\} $ with an energy gap of $\epsilon_{\rm{m}}$, see Fig. \ref{FIG2}a.
When performing gate operations, $\theta_{\rm{m}}$ is set to $\pi/2$, and the rotation axis is controlled by $\phi_{\rm{mw}}$.
In the laboratory frame, the $\vert 0''\rangle$ state is observed as an ideal Rabi oscillation at clock-frequency $\omega_{\rm{m}}/2\pi$, which starts in the $\vert 0\rangle$ state.
(The $\vert 1''\rangle$ state is a $\pi$ out of phase Rabi oscillation with $\vert 0''\rangle$, and starts in the $\vert 1\rangle$ state.)
The protected basis states coincide with the bare qubit states at every interval of $2\pi/\omega_{\rm{m}}$ - when this readout matching condition is met, we can readout the CCD-protected qubit state via energy-selective spin readout.
In other words, to measure
in the protected qubit basis the total length of the MW-pulse $t$ is set
such that $\omega_{\rm{m}} t=2\pi n$ ($n$ is an integer), when the protected qubit and bare-qubit bases are matched, $\vert\langle 1\vert 1''\rangle\vert^2=\vert\langle 0\vert0''\rangle\vert^2=1$.
We note that for $t=0$ the bases of the laboratory and the second rotating frames are the same, meaning that the CCD-protected qubit state can be initialized in the conventional, energy-selective manner.
When, on the other hand, $\omega_{\rm{m}} t=(2n+1)\pi$ the bases are inverted, $\vert\langle 1\vert 0''\rangle\vert^2=\vert\langle 0\vert 1''\rangle\vert^2=1$.
More generally, the total length of the MW pulse
selects the projection of the protected qubit basis to be measured, as depicted
in Fig.~\ref{FIG2}b.
In the following experiments, we insert idle pulses so that the total pulse time becomes an integer multiple of $2\pi/\omega_{\rm{m}}$ and realize the readout-match condition.
We summarize the different CCD drive conditions in Fig~\ref{FIG2}c.

We now show the experimental results of the basic operations in the CCD-protected basis.
In these experiments, we use $\epsilon_{\rm{m}} = \omega_{\rm{m}} /4 =2\pi\times235$ kHz and evaluate the decay by fitting to an exponentially decaying sinusoidal model, where the effect due to the counter-rotating term is neglected (see also Fig.~\ref{FIG4}b).
First, we demonstrate a Rabi oscillation between the protected qubit states, or CCD-Rabi oscillation, using the experiment shown in  Fig.~\ref{FIG3}a.
After initialization of the spin
state, we apply a first pulse, where $\theta_{\rm{m}} = \pi/2$ and
$\phi_{\rm{mw}}=0$ for a duration  $t_c$. This pulse
executes the $y$-rotation at frequency $\epsilon_{\rm{m}}/2\pi$, yielding a Rabi
oscillation in the protected basis. A subsequent idling-pulse with $\theta_{\rm{m}} = 0$ and $\phi_{\rm{mw}}=0$ is used to tune the total sequence time to the readout matching condition.
We find the decay time of the CCD-Rabi oscillation
$T_{\rm{CCD-Rabi}}$ = 54.9 $\mu$s and the quality factor $Q=25$.
Here, we define $Q$ by the $\pi$ rotation time over the Rabi decay time.
We furthermore demonstrate two-axis control of the protected qubit by applying two consecutive $\pi/2$ pulses with the MW phase $\phi_{\rm{mw}}$ of the second pulse varied.
As expected, the spin-up probability oscillates sinusoidally as a
function of $\phi_{\rm{mw}}$ with a period of $2\pi$ (Fig.~\ref{FIG3}b).

Next, we evaluate the coherence time of the CCD-protected qubit using
the experiment shown in Fig.~\ref{FIG3}c. First,
a $\pi/2$-pulse is applied. Since we set $\epsilon_{\rm{m}}=\omega_{\rm{m}}/4$,
the readout-indicator rotates by $2\pi$.
Then we apply the idling pulse for a duration $t_c$, which
corresponds to the free time evolution of a Ramsey interference
experiment, followed by the second $\pi/2$-pulse and another idling
pulse for readout matching. The CCD-Ramsey coherence measurement
shows a $z$-rotation with angular frequency $\epsilon_{\rm{m}}$ and a decay time of  $T_{\rm{CCD-Ramsey}}$ = 40.7 $\mu$s, which is more than two orders of magnitude longer than the bare $T_2^*$ measured.

\subsection{Evaluation of the gate fidelity}

\begin{figure*}[htbp]
    \includegraphics[width=140mm]{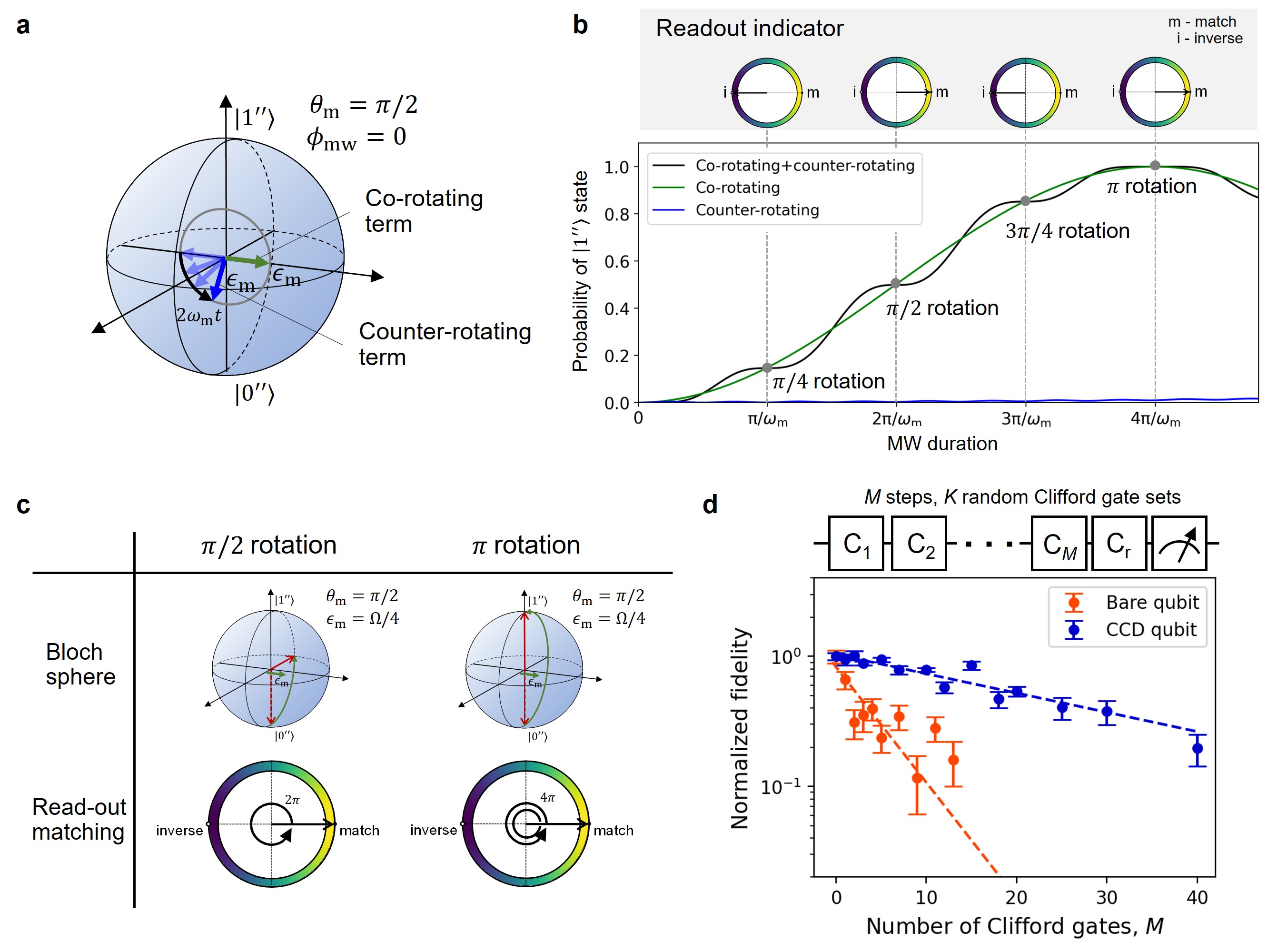}
    \caption{{\bf Gate operations and randomized benchmarking measurement.}
            {\bf a} Schematic illustration of co-rotating (CCD drive) term and
        counter-rotating term on the Bloch sphere. The Hamiltonian vector of
        the co-rotating term is constant in time, while that of the
        counter-rotating term rotates at angular frequency $2\omega_{\rm{m}}$.
            {\bf b} Numerical calculation of the probability $\vert 1''\rangle$ state due
        to the co-rotating and the counter-rotating term for $\epsilon_{\rm{m}} = \omega_{\rm{m}}/4$. The readout indicators are shown on the top.
        The rotation caused by the counter-rotating term is approximately zero at the period of $\pi/\omega_{\rm{m}}$.
            {\bf c} $\pi/2$ and $\pi$ rotations constituting Clifford gates. As we choose $\epsilon_{\rm{m}} = \omega_{\rm{m}}/4$, the indicator rotates $2\pi$ and $4\pi$ during the $\pi/2$ and $\pi$ rotation respectively, indicating that the readout condition is satisfied.
            {\bf d} Randomized benchmarking results for the bare and CCD-protected qubits. The sequence is repeated for $K=15$ random Clifford gate sets. The Clifford gates are composed of $\pi/2$ and $\pi$ pulses with durations of 1.06 $\mu$s and 2.13 $\mu$s, respectively. The sequence fidelities decay exponentially as a function of the number of Clifford gates. The gate infidelity is improved by a factor of 5 for the CCD-protected qubit.
        The average gate fidelity of the bare qubit is 95$\pm1\%$. The CCD spin control achieves a higher fidelity of 99.1$\pm 0.1\%$.
    }
    \label{FIG4}
\end{figure*}

In this section, we will evaluate the gate fidelity of the protected qubits. But first, we explain the choice of $\epsilon_{\rm{m}}=\omega_{\rm{m}}/4$ for gate-operations.
For simplicity, in the expression for the Hamiltonian in the protected frame $H_{\rm{rot}}^{(2)}$ given in Eq. \eqref{eq-004}, we have neglected a $2\omega_{\rm{m}}$ counter-rotating term.
In discussions of CCD, this is typically justified since the rotating wave approximation (RWA) holds for $2\omega_{\rm{m}} \gg \varepsilon_{\rm{m}}$.
However, RWA would require a very slow gate-speed, or otherwise the counter-rotating term is a potential source of gate-error.
To address this issue, we propose designing the gate time to be an integer multiple of the counter-rotating term, $2\omega_{\rm{m}} T_{\rm{gate}} = 2\pi n$, such that the counter-rotating term integrates to approximately zero without applying any idle pulse.
This minimizes the error arising from the counter-rotating term, whilst achieving a fast gate time.
To test this theoretically, a simulation of the control pulse is performed with and without the counter-rotating term included, see Fig.~\ref{FIG4}b, which indicates that the effect of the counter-rotating term cancels when the gate time is an integer multiple of $\pi/\omega_{\rm{m}}$.
Furthermore, setting $\epsilon_{\rm{m}} = \omega_{\rm{m}}/4$ enables the spin rotation in $\pi/4$ increments so that we can construct important quantum logic gates, such as T gate and Pauli gates.

Finally, we evaluate the CCD control fidelity via randomized benchmarking on Clifford gates~\cite{knill2008randomized,muhonen2015quantifying}, which is a well-known way to measure the gate fidelity in a robust manner against state preparation and measurement (SPAM) errors.
In this protocol, we apply $M$ random Clifford gates and a recovery Clifford gate.
Each Clifford gate is decomposed into $\pi$ and $\pm\pi/2$ rotations around the $x$- or $y$-axis except for the identity gate, which is implemented by no pulse (skipping).
In our experiment, we set $\epsilon_{\rm{m}}=\omega_{\rm{m}}/4 =2\pi\times235$ kHz, resulting in gate durations of 2.13 $\mu$s for $\pi$ rotations and 1.06 $\mu$s for $\pm\pi$/2 rotations.
The recovery Clifford gate is chosen such that the final target state is the spin-up or spin-down state. We calculate the difference in spin-up probability between the two cases, and evaluate the averaged gate fidelity from its decay.
To obtain one data point, the sequence is repeated for $K$ = 15 different random choices, and the sequence length $M$ is increased until the amplitude sufficiently decays, following a criterion similar to that used in refs.~\cite{muhonen2015quantifying,noiri2022fast}.
Note that for these Clifford gates, the readout matching conditions are automatically satisfied (Fig.~\ref{FIG4}c), so there is no need to add idle pulses.
Figure~\ref{FIG4}d shows the measurement result, where comparing the gate fidelities of the bare qubit and the CCD-protected qubit.
From a fit to an exponential curve $(2F_c-1)^M$, where $F_c$ is the average Clifford gate fidelity, we estimate the average single gate fidelity $F_c$ using the relation of $F_c^{\rm{single}} = 1-(1-F_c)/1.875$~\cite{muhonen2015quantifying}.
We extract $F_c^{\rm{single}}$ of 95$\pm1\%$ and 99.1$\pm 0.1 \%$ for the bare and the CCD-protected qubit, respectively.
This result demonstrates that our CCD spin control significantly improves the gate fidelity, reaching the critical threshold required for the surface code error correction~\cite{fowler2012surface}.

\section{Discussion}

In summary, a CCD control scheme has been applied to
the ESR of a single qubit in a natural Si-MOS
quantum dot. By dressing the spin with a phase-modulated drive which rotates the
spin about two axes, the spin is dynamically decoupled from both low frequency
errors in the detuning and the Rabi frequency.
A protocol to rapidly control the CCD-protected spin is proposed and demonstrated.
This protocol avoids the use of RWA to prevent the significant reduction in the gate speed, while protecting a qubit in the double dressed state, resulting in a significant improvement in the gate fidelity from 95$\%$ to 99$\%$ despite the naturally abundant $^{29}$Si environment.
We would like to emphasize that our experiment is achieved without any feedback or continual calibration method~\cite{shulman2014suppressing,nakajima2020coherence,berritta2024real}, demonstrating the robustness and reliability of our approach.

We evaluate the single-qubit gate fidelity achieved using the CCD protocol via randomized benchmarking. In our experiment, the coherence time under CCD driving, $T_{\rm{CCD-Rabi}}$ was measured to be 54.9 $\mu$s, while the average single-qubit gate duration $t_{\rm{ave}}$ used in randomized benchmarking was 1.23 $\mu$s.
Under the assumption that coherence time is the dominant source of infidelity, the average gate fidelity can be approximated by the expression~\cite{stano2022review}:
$F = \frac{1}{2}[1+\exp(-t_{\rm{ave}}/T_{\rm{CCD-Rabi}})]$.
This estimation yields the average single-gate fidelity of 98.9$\%$, consistent with the observed value of 99.1$\%$.
This result suggests that increasing the Rabi frequency to shorten the gate duration is a promising strategy to further improve the fidelity.
Agreement with the above approximation implies that the effect of non-Markovian noise~\cite{knill2008randomized,fogarty2015nonexponential,tanttu2024assessment} on our randomized benchmarking result is minor.
However, this remains a subject for future investigation, as the CCD method leverages the non-Markovian nature of the noise to enhance gate fidelities.
While the reported value may deviate from the true fidelity, we emphasize that the main point of this study is the large improvement in the quality of CCD based control compared to regular Rabi pulses in the noisy device.

For the noisy qubit studied here, there is a strong benefit to using CCD.
The robustness of CCD to low frequency noise arises from a Rabi oscillation that is insensitive to detuning and Rabi frequency, and this property is highly desirable for global control schemes with some variability between qubits~\cite{vahapoglu2021single,vahapoglu2022coherent}.
Furthermore, during a quantum algorithm, qubits will spend most of the time waiting~\cite{fowler2012surface,anastasiou2024tetris,long2024layering}, and CCD looks promising for idling operations.

The CCD protocol offers advantages for global control in large-scale qubit arrays, as the robustness against detuning errors also reduces the need for precise phase tracking of individual qubits. Previous studies have demonstrated that 3D microwave dielectric resonators can generate global fields suitable for coherent spin control~\cite{vahapoglu2021single,vahapoglu2022coherent}, which could be a promising direction for implementing CCD in scalable architectures. Prior studies have demonstrated enhanced robustness in two-qubit operations using SMART protocol~\cite{hansen2024entangling}, and similar control scheme could be used to implement robust CCD-based two-qubit gates.

\section*{Methods}

\subsection*{Device fabrication}
The device is fabricated using standard CMOS fabrication technique, including a self-aligned gate process~\cite{lee2020enhancing}.
Figures~\ref{FIG1}a and b. show a scanning electron micrograph and a cross-sectional schematic of the sample used in this study.
On a 145 nm thick buried oxide layer, the T-shaped natural silicon channel is fabricated by the STI process.
For the gate fabrication, we use the self-aligned double patterning processes in two lateral directions to reduce the gate pitch~\cite{kuno2025single}.
The plunger gates in the second layer (SG) are formed by a self-aligned process using the barrier gates in the first layer (FG).
A 15 nm layer of SiO$_2$ is then deposited, followed by an 80 nm layer of phosphorus-doped poly-Si.
The third layer (TG) gates for SET are then formed again by the self-alignment process using the SG plunger gates.
The aluminum microstrip line (600 nm wide and 500 nm thick) is placed 300 nm above the QD array for ESR.

\subsection*{Experimental setup}
The sample is cooled down in a dilution refrigerator (Oxford Instruments Proteox) with a superconducting magnet.
All measurements are performed at base temperature of 10 mK and magnetic field of 825 mT.
The low-noise voltage generator (Qdevil QDAC-II) is used to apply gate voltages through low pass filters.
The SET current is amplified by a current-voltage amplifier (Basel SP983c) and then monitored with a digitizer (Zurich Instruments UHFLI).
For spin manipulation, a.c. magnetic field is generated by a MW source (Keysight PSG E8257D) and modulated by an AWG (Zurich Instruments HDAWG) with an I/Q mixer (Analog Devices HMC8192LG).
Supplementary Note 6 provides the experiment wirings.
Note that, in our experiment, the electron temperature is relatively high, around 240 mK, and we employ Elzerman readout. This leads to large SPAM errors and reduced visibility in the experimental data.

\section*{Acknowledgement}
This work was supported by JST Moonshot R$\&$D Grant Number JPMJMS2065, Grants-in-Aid for Scientific Research grant numbers JP23H05455 and JP23K17327, and JST PRESTO Grant Number JPMJPR21BA.

\section*{Author contributions}
T.Ku. performed the experiment and analyzed the data.
T.Ku., T.U., A.J.R., N.M., R.M., T.N, J.Y., T.Ko., S.S., D.H. and R.T. discussed the results.
T.Ku., T.U. and N.K contributed to the measurement setup.
N.L., I.Y., T.M., D.H. and R.T. designed and fabricated the device.
T.Ku., T.U. and A.J.R. performed calculations.
T.Ku. and A.J.R. wrote the manuscript with input from all co-authors.
H.M. supervised the project.

\section*{Data availability}
Data is available from the corresponding author upon reasonable request.

\section*{Competing interests}
The authors declare no competing interests.

\clearpage

\renewcommand{\bibsection}{\section*{References}}
\bibliographystyle{naturemag}
\bibliography{refs}

\clearpage
\newpage
\onecolumngrid

\section*{Supplementary Note 1: Bare qubit properties}

\begin{figure}[htbp]
    \includegraphics[width=\columnwidth]{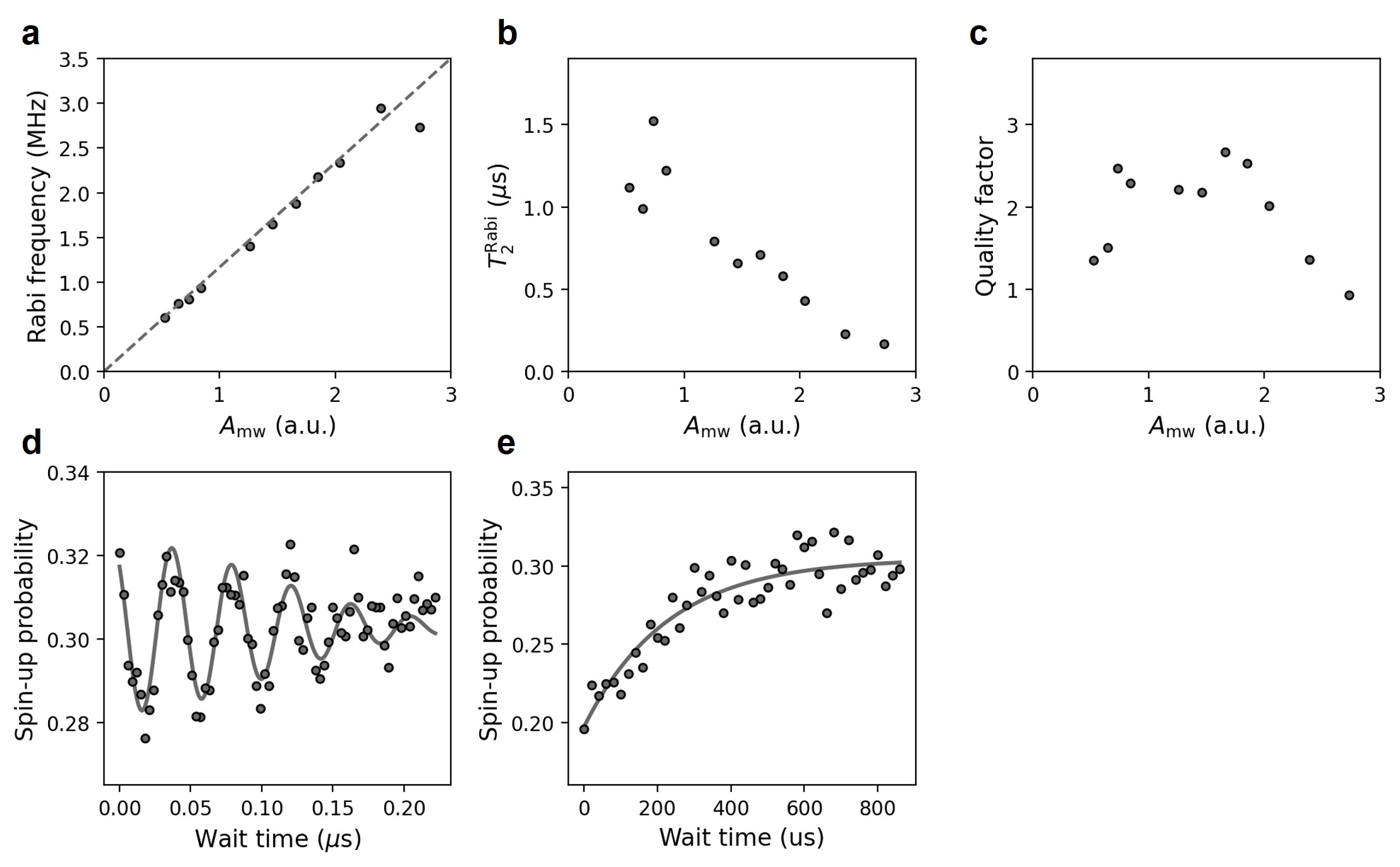}
    \caption{{\bf Bare qubit properties.}
        {\bf a} Micorwave amplitude $A_{\rm{mw}}$ dependence of Rabi frequency. The dotted line is a linear fitting for $A_{\rm{mw}}\leq 2.4$. Rabi oscillations above 3 MHz cannot be obtained due to the strong damping.
        {\bf b} $A_{\rm{mw}}$ dependence of Rabi decay time $T_2^{\rm{Rabi}}$.
        {\bf c} $A_{\rm{mw}}$ dependence of quality factor. Since $T_2^{\rm{Rabi}}$ rapidly decreases as $A_{\rm{mw}}$ is increased, the quality factor is limited to 2.7.
        {\bf d} Ramsey sequence. The relative phase of the second pulse is artificially modulated at 25 MHz to improve the fit. The coherence time $T_2^{\rm{*}}$ is 143 ns.
        {\bf e} Spin echo result. A spin echo sequence improves the coherence time and we obtain $T_2^{\rm{echo}} = 230$ $\mu$s. $T_{\rm{2}}^{\rm{echo}}$ is more than 1000 times longer than $T_{\rm{2}}^{\rm{*}}$, suggesting that the source of dephasing is low frequency noise compared with the timescale of the electron spin dynamics.
    These results are consistent with those of previous measurements~\cite{kuno2024concatenated}.
    }
    \label{Barequbit}
\end{figure}

\newpage
\section*{Supplementary Note 2: Concatenated continuous driving}
This section explains the theory of CCD in more detail.
There have been several works on coherence protection using CCD~\cite{wang2020coherence,cohen2017continuous,stark2017narrow,ramsay2023coherence}.
Two methods of CCD are typically studied: amplitude-modulated CCD and phase-modulated CCD.
When IQ (in-phase and quadrature) signals are modulated with an arbitrary waveform generator (AWG), the frequency stability is better than its power stability.
Therefore, we adopt phase-modulated CCD and the system Hamiltonian in the laboratory frame can be expressed as

\begin{align}
     & H_{\rm{lab}}(t)=\frac{\hbar(\omega_{\rm{mw}}+\delta)}{2} \sigma_z
    + \hbar\Omega \cos\left[\omega_{\rm{mw}}t+\phi_{\rm{mw}} - \frac{2\epsilon_{\rm{m}}}{\omega_{\rm{m}}}\sin(\omega_{\rm{m}}t-\theta_{\rm{m}})\right]\sigma_x.
\end{align}

In the phase-modulated CCD protocol, it is convenient to take the
first rotating frame of the time-dependent Hamiltonian defined by
$H_{0}^{(1)} = \frac{\hbar}{2}\frac{d}{dt}\Phi_{\rm{CCD}}\sigma_z$, where
$\Phi_{\rm{CCD}}=\omega_{\rm{mw}}t-\frac{2\epsilon_{\rm{m}}}{\omega_{\rm{m}}}\sin(\omega_{\rm{m}}t-\theta_{\rm{m}})$.
Then the Hamiltonian of the first rotating frame, $H_{\rm{rot}}^{(1)}(t) = e^{i\Phi_{\rm{CCD}}\sigma_z/2}H_{\rm{lab}}e^{-i\Phi_{\rm{CCD}}\sigma_z/2} - H_{0}^{(1)}$ is described as,

\begin{align}
    H_{\rm{rot}}^{(1)}(t) & =\frac{\hbar\delta}{2}\sigma_z
    + \frac{\hbar\Omega}{2} \sigma_{\phi_{\rm{mw}}}
    + \hbar\epsilon_{\rm{m}}\cos(\omega_{\rm{m}}t-\theta_{\rm{m}})\sigma_z\label{sup-1strot}
    + \frac{\hbar\Omega}{2} \left[\cos(2\Phi_{\rm{CCD}}+\phi_{\rm{m}})\sigma_x - \sin(2\Phi_{\rm{CCD}}+\phi_{\rm{m}})\sigma_y\right],
\end{align}
where we define ${\sigma_{\phi_{\rm{mw}}}=\cos(\phi_{\rm{mw}})\sigma_x + \sin(\phi_{\rm{mw}})\sigma_y}$.
The second term represents the usual Rabi driving field, the third term arises from the MW phase-modulation, and the fourth is the counter-rotating term.
When the Rabi frequency is much smaller than the resonance frequency, i.e. $\Omega \ll \omega_{\rm{mw}}$, the rotating wave approximation (RWA) is valid and the counter-rotating term can be neglected.
This condition is easily achieved, because for a typical spin qubit in silicon  $\Omega$ is up to about 20 MHz  and $\omega_{\rm{m}}$ is in the order of ten GHz.

In the case of a freely evolving spin ($\Omega=\epsilon_{\rm{m}}=0$), a detuning error
mainly due to $^{29}$Si nuclear spins causes a phase-shift that grows linearly with both time and detuning, leading to a short $T_2^*$ time measured
via a Ramsey sequence. For a Rabi-drive ($\epsilon_{\rm{m}}=0$, $\Omega\neq0$), a detuning error tilts the Hamiltonian vector in the rotating frame,
resulting in a rotation error that grows linearly with time, but quadratically
with error in detuning, resulting in a longer $T^{\rm{Rabi}}_{2}$ time.

To illustrate the role of the third term, we further move to the second rotating frame defined by $H_0^{(2)}=\frac{\hbar\omega_{\rm{m}}}{2}\sigma_{\phi_{\rm{mw}}}$.
The Hamiltonian of the second rotating frame, $H_{\rm{rot}}^{(2)}(t) = e^{\frac{i\omega_{\rm{m}}\sigma_{\phi_{\rm{mw}}}}{2}t}H_{\rm{rot}}^{(1)}e^{\frac{-i\omega_{\rm{m}}\sigma_{\phi_{\rm{mw}}}}{2}t} - H_{0}^{(2)}$, is

\begin{align}
     & H_{\rm{rot}}^{(2)}(t)=\frac{\hbar(\Omega-\omega_{\rm{m}})}{2}\sigma_{\phi_{\rm{mw}}}
    +\frac{\hbar\epsilon_{\rm{m}}}{2} \left\{ \cos(\theta_{\rm{m}})\sigma_z+\sin(\theta_{\rm{m}})\sigma_{\phi_{\rm{mw}+\frac{\pi}{2}}}  \right\}
    +\frac{\hbar\epsilon_{\rm{m}}}{2} \left\{ \cos(2\omega_{\rm{m}} t-\theta_{\rm{m}})\sigma_z + \sin(2\omega_{\rm{m}} t-\theta_{\rm{m}})\sigma_{\phi_{\rm{mw}+\frac{\pi}{2}}} \right\},
\end{align}

where we set $\delta = 0$. Assuming the on-resonance condition, where we set $\omega_{\rm{m}}$ to the mean value of $\Omega$, the first term is considered to be the Rabi frequency error.
The second term is the CCD driving field, and the third term is the counter-rotating term.
The counter-rotating term can be neglected when the condition $\epsilon_{\rm{m}}\ll 2\omega_{\rm{m}} = 2\Omega$ is satisfied (RWA).
However, it is not desirable to set $\epsilon_{\rm{m}}$ so small because it slows down the CCD gate operation speed.
In this work, we propose that by setting the total pulse duration at $t = 2n\pi/\omega_{\rm{m}}$ ($n$ is an integer), we minimize the impact of the counter-rotating term.
The CCD-drive acts as an effective d.c. magnetic field oriented by
the microwave phases $\theta_{\rm{m}}$ and $\phi_{\rm{mw}}+\frac{\pi}{2}$ and is perpendicular
to the Rabi frequency error, reducing the impact of errors in the Rabi drive in a manner analogous to the case of the detuning error.
Furthermore, in this frame, the detuning error is expressed as a second-harmonic and is dynamically decoupled from the spin dynamics, further increasing the coherence time.

To readout the CCD-protected qubit, we consider
the relationship between the state vectors of the second rotating frame and the laboratory frame.
The state vectors of the second rotating frame $|\psi\rangle_{\rm{rot}}^{(2)}$ is described as,
\begin{align}
    |\psi\rangle_{\rm{rot}}^{(2)} & = \exp\left(iH_0^{(2)}t\right)\exp\left(i\int H_0^{(1)} dt\right)|\psi\rangle_{\rm{lab}}                                                         \\
                                  & = \exp\left(\frac{i\omega_{\rm{m}}\sigma_{\phi_{\rm{mw}}}}{2}t\right)\exp\left(\frac{i\Phi_{\rm{CCD}}\sigma_z}{2}\right)|\psi\rangle_{\rm{lab}}.
\end{align}
We note that the Pauli matrices of the first and second operators are defined in the laboratory and the first rotating frame, respectively.
Considering the readout in the laboratory frame,
the first operator $\exp\left(\frac{i\Phi_{\rm{CCD}}\sigma_z}{2}\right)$ only adds a phase to the state vector in the laboratory frame, which does not change the readout results.
On the other hand, the second operator $\exp\left(\frac{i\omega_{\rm{m}}\sigma_{\phi_{\rm{mw}}}}{2}t\right)$ changes the spin polarization in the laboratory frame.
Except for the addtional phase, the state vector of the second rotating frame matches that of the laboratory frame in a period of $2\pi/\omega_{\rm{m}} $.
In this work, we add the idle pulses such that the total pulse length is an integer multiple of $2\pi/\omega_{\rm{m}} $ for readout matching.

\newpage
\section*{Supplementary Note 3: Comparison of the bare qubit and the CCD-protected qubit}

In this section we discuss the impact of low frequency noise on the damping of Rabi oscillations, and compare the bare qubit and the CCD-protected qubit.
In particular, we focus on the magnetic noise caused by the nuclear spins from $^{29}$Si isotopes and treat it as a quasi-static Overhauser field~\cite{hanson2007spins}.

Consider the textbook Rabi problem of a spin-1/2, $H_0 = \frac{\hbar\omega_0}{2}\sigma_z$ driven near resonance by an a.c. drive of frequency $\omega/2\pi$, and Rabi frequency $\Omega/2\pi$. In the lab frame the drive Hamiltonian is:
\begin{equation}
    H_{\rm{Rabi}} = \hbar\Omega\cos{(\omega t)}\sigma_x.
\end{equation}

When we consider a spin with no decoherence, the final expectation value of spin-$z$ can be calculated versus the duration of the MW-pulse and the detuning from resonance $\Delta = \omega-\omega_0$, see Supplementary Figure~\ref{Rabichevron}a.
The color map displays the signature Rabi chevrons. These contours arise due to the effective Rabi frequency $\sqrt{\Omega^2+\Delta^2}$, and the width is set by the tilt in the rotation axis.
If the dominant source of noise is a fluctuation in the qubit frequency or the drive that is slow on the timescale of a single measurement run, the qubit acts as an ideal spin: for each measurement run, there is an uncertainty in the detuning or drive of the control pulse, resulting in a reversible error.
However, after many runs, the result averages over some variations in the rotation angle and axis leading to damping.
This is simulated in Supplementary Figures~\ref{Rabichevron}c-g, where the data of Supplementary Figure~\ref{Rabichevron}a is averaged over the detuning weighted by a normalised factor $ f(\Delta)\propto \exp{(-\frac{\Delta^2}{2\sigma^2})}$.
As we assume that the detuning error is caused by the nuclear spins from $^{29}$Si isotopes, the standard deviation $\sigma$ and coherence time $T_2^*$ have the relation $\sigma = \sqrt2/T_2^*$.
In this simulation, we calculate the 1 MHz Rabi oscillations for a typical silicon spin qubit with coherence time $T_2^*$ of 100 ns to 10 $\mu$s.
A shorter coherence time means a larger detuning error, and the Rabi oscillation rapidly decays.

The CCD drive uses the phase modulation to reduce the sensitivity to uncertainties in the detuning and drive.
The drive Hamiltonian for CCD is:

\begin{eqnarray}
    H_{\rm{CCD}} = \hbar\Omega\cos{\left[\omega t - \frac{2\epsilon_{\rm{m}}}{\omega_{\rm{m}}}\sin(\omega_{\rm{m}} t -\theta_{\rm{m}})\right]}\sigma_x, \nonumber \\
\end{eqnarray}
with the resonance drive condition given by $\omega_{\rm{m}} = \Omega$.
The Rabi oscillation period of the encoded qubit is set by the modulation frequency $\omega_{\rm{m}}$ (a time-frequency), rather than the relatively noisy Rabi drive set by an MW-power.
Supplementary Figure~\ref{Rabichevron}b presents a Rabi chevron for the CCD protocol.
The chevron pattern has been straightened out, expressing a reduced sensitivity to the detuning errors.
The Rabi oscillations with the CCD protocol are shown in Supplementary Figures~\ref{Rabichevron}h-l.
The decay observed in the bare qubit is suppressed, and the Rabi oscillations continue for a long time.
Note that the amplitude of the Rabi oscillations cannot be recovered even if the CCD protocol is employed.
This is because a qubit that has deviated from the on-resonance condition given by the ladder-like shape Rabi chevron cannot be driven and can be considered a state preparation error in the context of the CCD-protected qubit.

\begin{figure*}[h]
    \includegraphics[width=\columnwidth]{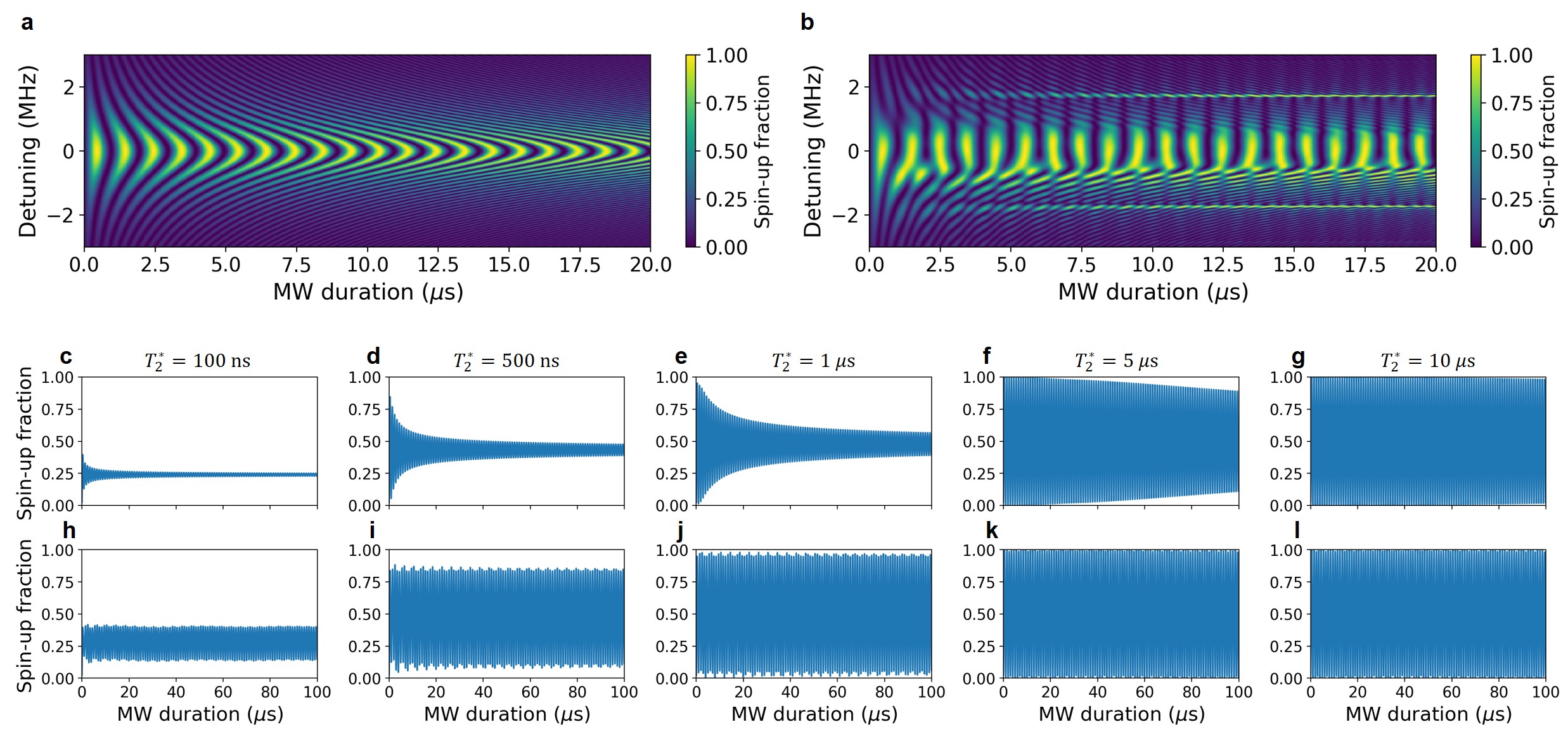}
    \caption{{ \bf Numerical simulation comparing the bare qubit and the CCD-protected qubit with the Rabi frequency $\Omega/2\pi$ = 1 MHz.}
            {\bf a} Rabi chevrons for bare qubit. The curvature of the contours arises from the effective Rabi frequency, $\Omega_{\rm{eff}}=\sqrt{\Omega^2+\Delta^2}$, and indicates that the rotation angle is sensitive to detuning.
            {\bf b} Rabi chevrons for a CCD-protected qubit, where $\epsilon_{\rm{m}} = \omega_{\rm{m}}/4$. A ladder-like pattern is observed, indicating an enhanced insensitivity of the rotation angle to the detuning.
            {\bf c-g} Rabi oscillations averaged over detuning with a weighting factor $ f(\Delta)\propto \exp{(-\frac{\Delta^2}{2\sigma^2})}$, where $\sigma=\sqrt2/T_2^*$.
            {\bf h-l} Rabi oscillations for CCD protocol with the same coherence time $T_2^*$ as the bare qubit.
    }
    \label{Rabichevron}
\end{figure*}

\clearpage

\section*{Supplementary Note 4: Robustness to detuning and Rabi frequency errors}

\begin{figure}[htbp]
    \includegraphics[width=\columnwidth]{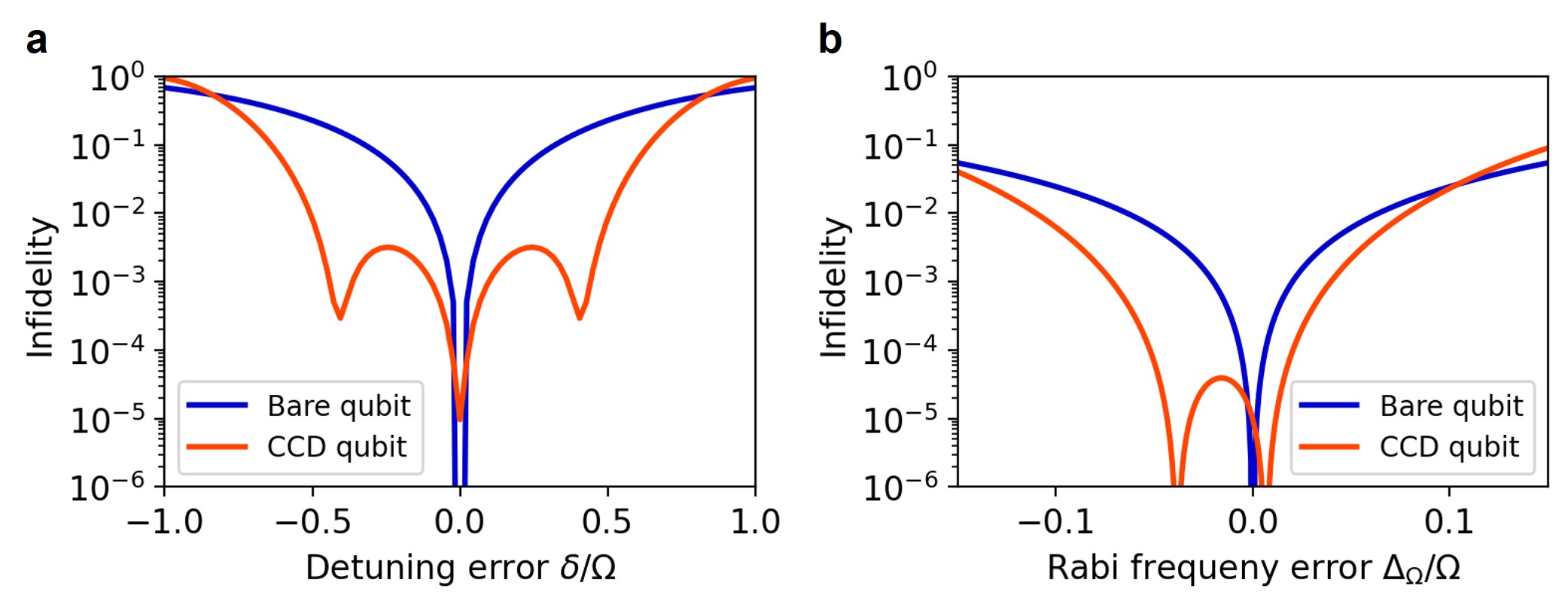}
    \caption{{\bf Numerical simulaiotns of the $\pi$ rotation infidelity.}
            {\bf a} Detuning error.
            {\bf b} Rabi frequency error.
        Modulation parameters are $\epsilon_{\rm{m}}=\omega_{\rm{m}}/4$, $\theta_{\rm{m}}=\pi/2$, and $\phi_{\rm{mw}}=0$, the same as those employed in the experiment.}
    \label{Robustness}
\end{figure}

One key advantage of the CCD protocol is its inherent robustness against both detuning and Rabi frequency errors.
To evaluate this robustness, we perform numerical simulations of the infidelity of $\pi$ rotations and compare them with the bare qubit property.
Supplementary Figure~\ref{Robustness} shows infidelity of $\pi$ rotations, where the modulation parameters are the same as those used in the experiment described in the main text.
In this simulation, we adopt a quasi-static noise model, in which the errors remain constant during the $\pi$ rotation.
The Rabi frequency error is defined as $\Delta_\Omega = \Omega-\omega_{\rm{m}}$, where we set $\omega_{\rm{m}}$ to be the mean value of $\Omega$.
The results demonstrate enhanced robustness of the CCD protocol over the bare qubit for both types of errors.
Compared to a bare qubit, the CCD protocol exhibits significantly enhanced robustness against both types of errors.
Although the bare qubit achieves slightly lower infidelity at zero error, this is due to residual counter-rotating terms in the CCD protocol.
Importantly, the CCD protocol maintains high fidelity while offering substantial robustness across a broad range of errors.
The CCD protocol can reduce the need for precise calibration or feedback in practical implementations.

\clearpage
\newpage
\section*{Supplementary Note 5: Gate duration and infidelity in the CCD protocol}

\begin{figure}[htbp]
    \includegraphics[width=140mm]{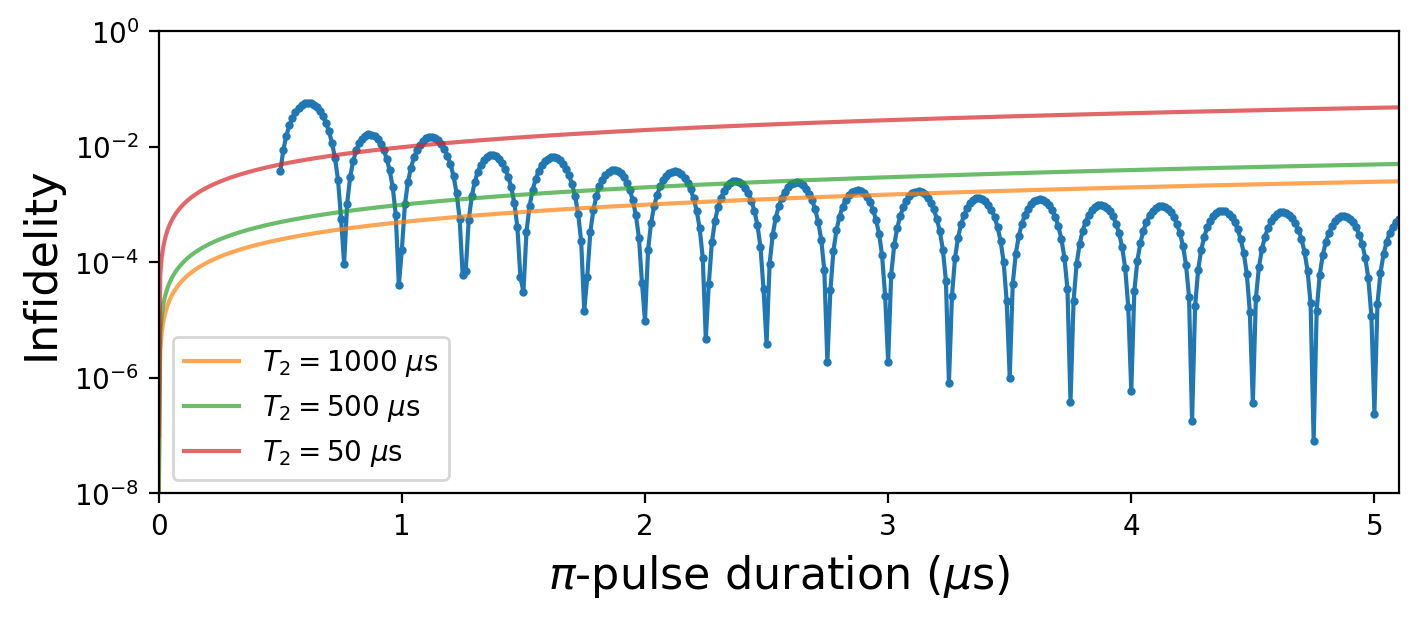}
    \caption{{\bf Numerical simulation of a $\pi$-pulse infidelity in the CCD protocol as a function of pulse duration.}
        We use Rabi frequency $\Omega/2\pi$ = 1 MHz and, CCD modulation parameters of $\omega_{\rm{m}}=\Omega$ and $\phi_{\rm{mw}}=0$.
        $\pi$-pulse duration $t_{\pi}$ in the CCD protocol is determined by $\epsilon_{\rm{m}}$, following relation $t_{\pi}=\pi/\epsilon_{\rm{m}}$.
        The solid blue lines shows the infidelity of the $\pi$-pulse, while the semi-transparent lines show the infidelity due to decoherence for comparison.}
    \label{Duration}
\end{figure}

The gate duration of a CCD-protected qubit is determined by the modulation parameter $\epsilon_{\rm{m}}$.
Supplementary Figure~\ref{Duration}a shows the infidelity of a $\pi$-pulse plotted against its duration, $t_{\pi}=\pi/\epsilon_{\rm{m}}=n\pi/\omega_{\rm{m}}$, where the Rabi frequency $\Omega/2\pi$ is set to 1 MHz and the modulation frequency $\omega_{\rm{m}}$ is matched to the Rabi frequency, $\omega_{\rm{m}}=\Omega$.
Initially, as the $\pi$-pulse duration increases, the infidelity displays a damped oscillation.
This is because in the protected qubit frame, which rotates in sync with Rabi oscillation, there also exists an undesired counter-rotating term at frequency $2\omega_{\rm{m}}$.
The envelope decays, since as $\epsilon_{\rm{m}}/\omega_{\rm{m}}$  approaches zero, the effect of the counter-rotating frame integrates to zero in a long-time limit.
In addition, when $n=\omega_{\rm{m}}/\epsilon_{\rm{m}}$ takes half-integer values, the effect of the counter-rotating term integrates to almost zero, giving rise to the observed oscillations.
In the longer gate time regime, decoherence starts to limit the fidelity.
In this study, we adopted $\epsilon_{\rm{m}}$ such that $\epsilon_{\rm{m}} =\omega_{\rm{m}}/4$ (corresponding to 2 $\mu$s in Fig.~\ref{Duration}) to exploit the cancellation effect.
An additional benefit of choosing $\epsilon_{\rm{m}} =\omega_{\rm{m}}/4$ is that the readout timing aligns naturally during Clifford gate operations, without requiring extra idle pulses.

\clearpage
\newpage
\section*{Supplementary Note 6: Experimental setup}

\begin{figure}[htbp]
    \includegraphics[width=140mm]{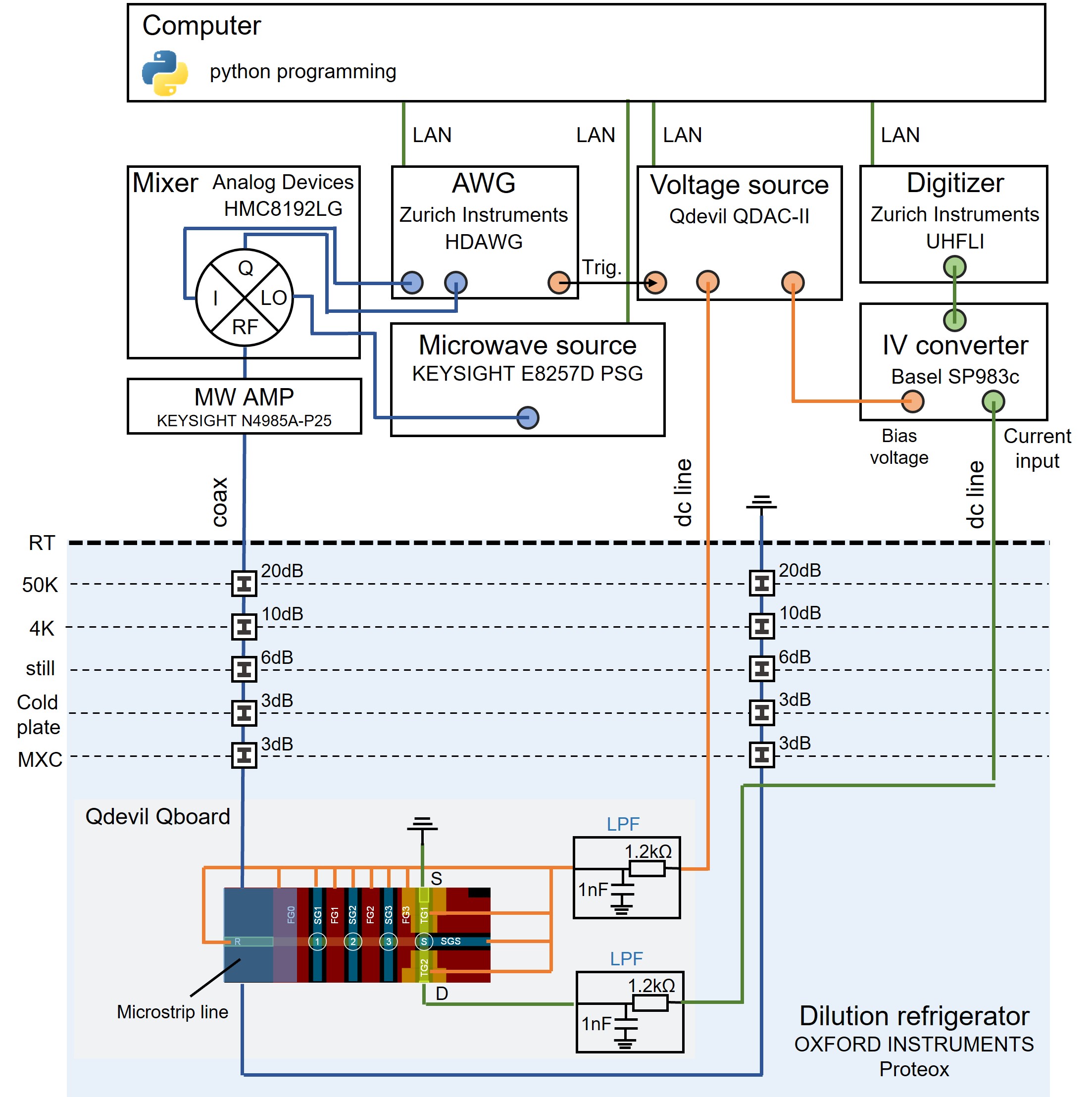}
    \caption{{\bf Schematic of experimental setup.}
        The device is wire-bonded to the PCB and mounted on the cryogenic sample holder (Qdevil Qboard).
        The sample holder has low pass filters (LPFs), and gate voltages are applied through these LPFs.
        We apply DC and pulse voltages to each gate using a low-noise voltage generator (Qdevil QDAC-II).
        The microwave for spin manipulation is generated by a microwave source (Keysight PSG E8257D) and modulated by the mixer (Analog Devices HMC8192LG).
        I/Q signals are generated by an AWG (Zurich Instruments HDAWG).
        After the microwave is amplified by the amplifier (Keysight N4985A-P25), it is input to the refrigerator and transmitted by the coaxial cable to the microstrip line.
        The coaxial cables have attenuators at different temperature stages to reduce the heat load.
        SET current is converted to a voltage signal with an IV converter (Basel SP983c) and monitored by a digitizer (Zurich Instruments UHFLI).
    }
    \label{ExperimentalSetup}
\end{figure}
\clearpage

\end{document}